\documentstyle{article}
\begin{document}
\title{Elastic fields of stationary and moving dislocations
in three dimensional finite samples}
\author{{\sc Rodrigo Arias} and {\sc Fernando Lund} \\ 
Departamento de F\'\i sica, Facultad de Ciencias F\'\i sicas y  
Matem\'aticas \\ Universidad de Chile, Casilla 487-3, Santiago, Chile}
\date{\today}
\maketitle
\newcommand{\beq}{\begin{equation}}
\newcommand{\eeq}{\end{equation}}
\newcommand{\bea}{\begin{eqnarray}}
\newcommand{\eea}{\end{eqnarray}}

\begin{abstract}
Integral expressions are determined for the elastic displacement and
stress fields due to stationary or moving dislocation loops in three 
dimensional, not necessarily isotropic, finite
samples. 
A line integral representation is found for the stress field, thus 
satisfying the expectation that stresses should depend on the location of 
the dislocation loop, but not on the location of surfaces bounded by such 
loops that are devoid of physical significance.

In the stationary case the line
integral representation involves 
a ``vector potential'' that depends on the specific geometry of the sample, 
through its Green's function:
a specific combination of derivatives of the elastic stress produced by the
Green's function appropriate for the sample is divergenceless, so it is
the curl of this ``vector potential''.
This
 ``vector potential''  is explicitely determined 
for an isotropic half space
and for a thin plate. Earlier specific results in these geometries are 
recovered as special cases.

In the non stationary case a line integral representation can be obtained
for the time derivative of the stress field. This, combined with
the static result, assures a line integral representation for the time
dependent stress field.
\end{abstract}

\section{Introduction}
A general formula for
the displacement and stress fields generated by a dislocation loop
undergoing arbitrary motion in an infinite medium
was obtained some years ago by
Mura (1963). In this formulation, the stress field is written
as a convolution of
the medium's impulse
response with a source localized along the dislocation loop
(i.e. independent of the loop's slip plane).
These formulae do not apply to the case of finite samples since
they are derived assuming homogeneity in space, i.e. a dependence
of the Green's function in $\vec{x}-\vec{x}'$, a fact that is
no longer true in a finite sample.
The purpose of this paper is to
present a generalization of these formulae to
finite samples, as well as
explicit forms for the geometries of a half space and
thin plate  in an isotropic elastic medium.

Motivation for this work comes partly from an ongoing project that
attempts to understand recently observed dynamic instabilities for
cracks in thin plates (Sharon, Gross and Fineberg, 1996, Boudet,
Ciliberto and Steinberg, 1996) in terms of the interaction of the
plate's oscillations with the crack tip regarded as a continuous
distribution of infinitesimally small dislocations (Lund, 1996).
An intuitive thought behind this work was that the stress fields produced by
stationary or moving dislocation loops in a finite sample
should not depend on the slip planes chosen, and that indeed
the stresses will be continuous through the slip planes,
as it is the case in an infinite medium. This
is rigorously shown here: the expressions
found for the stress fields involve only line integrals over
the dislocation loops. In addition, Gosling and Willis (1994) have 
emphasized the importance of dealing with free surface effects in order to 
understand the strain relaxation within thin strained layers over a 
substrate such as are employed in the manufacture of high technology 
devices. They developed a line integral representation for the stresses due 
to an arbitrary dislocation in an isotropic half-space, in which the 
integrand can be constructed by an explicitely given algorithm. The 
resulting integral can be explicitely evaluated for a dislocation half-line. 
Line integral representations are computationaly more efficient than the 
more easily derived surface integral representations. 

 	In order to find the explicit
form of the line integrals appearing in the stress fields 
when the loops are stationary, one
should calculate a ``vector potential'', that is, a vector quantity  
whose curl gives a combination of derivatives of
 the inhomogeneous portion of the Green's function,  
appropriate to the finite
sample in question. We have explicitely done this for two examples:
an isotropic half space and a thin plate.
In addition to the calculation of the ``vector potential''  
in these two geometries
we have explicitely worked out the case of a stationary screw dislocation
perpendicular to the free surfaces. These two examples can also be  
obtained
independently with the method of Eshelby and Stroh (1951), thus  
providing a check on the algebra.
In a non stationary
situation the displacement velocity field can be written in
terms of line integrals,
and from this one deduces that the
time derivative of the stress field can be written
in terms of them too. One finally deduces that the
time dependent stress field can be written in terms of line
integrals since one knows that the initial condition
(stationary case) and its temporal derivative can both be
written in terms of line integrals.

\section{Displacement fields due to dislocations in finite samples}

Here we derive a formula for the elastic displacement field 
produced in a finite sample 
by a moving dislocation loop, that is written as an
integration over succesive slip planes.

We consider small displacements $U_{m}(x,t)$ in
an homogeneous elastic medium of density $\rho$ and elastic constants
$C_{ijkl}$. A general formula for
the displacement, velocity and stress fields generated by a dislocation loop
undergoing arbitrary motion in an infinite medium
was obtained some years ago by
Mura (1963). 
The displacement field is written as an integration over succesive
slip planes, where the displacement field has a discontinuity
given by the Burgers vector $b_{i}$.
The velocity and stress fields are written
as a convolution of
the medium's impulse
response with a source localized along the dislocation loop,
i.e. these fields are independent of the choice of slip planes.
In this section Mura's formula for the displacement field is 
generalized to a finite sample.

The particle displacement obeys the wave equation:
\begin{equation}
\label{eq:wave}
\rho \frac{\partial^{2} U_{i}}{\partial t^{'2}}(x',t') - C_{ijkl}
\frac{\partial^{2} U_{k}}{\partial x_{l}' \partial x_{j}'}(x',t')=0 \: ,
\end{equation}
and normal stress-free boundary conditions at the surfaces of the
sample:
\begin{equation}
\label{bc}
C_{ijkl}\frac{\partial U_{k}}{\partial x_{l}'} (x_{S}',t)n_{j}(x_{S}')
=0 \: ,
\end{equation}
with $x_{S}'$ a point on the surface and $n_{j}(x_{S}')$ its 
outwardly-pointing
normal vector. The medium's impulse response is given by a
Green's function $G_{im}(x,x';t-t')$, which is the displacement 
in direction $(i)$ evaluated at $(x,t)$ produced by a localized impulse 
force
in direction $(m)$ applied at position $x'$ and time $t'$. 
It is the solution of:
\begin{equation}
\rho \frac{\partial^{2} G_{im}}{\partial t^{2}}(x,x';t-t') -
C_{ijkl}\frac{\partial^{2} G_{km}}{\partial x_{l} \partial x_{j}}
(x,x';t-t') = \delta_{im}\delta(x-x')\delta(t-t')
\: .
\end{equation}
We have explicitely exhibited the $x$ and $x'$ dependence because the 
order in which they apear is important.
If one defines as $\sigma_{ij}^{Gm}(x,x';t-t')$ the elastic stress
associated with the Green's function:
\beq
\sigma_{ij}^{Gm}(x,x';t-t') \equiv 
C_{ijkl}\frac{\partial }{\partial x_{k}} G_{lm}(x,x';t-t') \: ,
\eeq
 the free surface boundary condition reads:
\begin{equation}
\label{eq:bcg}
\sigma_{ij}^{Gm}(x_{S},x';t-t')n_{j}(x_{S})=0 \: .
\end{equation}
Note that, since the medium is bounded, there is homogeneity in time but
not in space.

In order to find the displacement field due to a moving dislocation
we start, following Mura (1963), with the following identity that holds 
due to a
symmetry of the elastic constants tensor ($C_{ijkl}=C_{klij}$):
\begin{equation}
C_{ijkl}\frac{\partial U_{k}}{\partial x_{l}'}(x',t')
\frac{\partial G_{mi}}{\partial x_{j}'}(x,x';t-t') = C_{ijkl}
\frac{\partial U_{i}}{\partial x_{j}'}(x',t')\frac{
\partial G_{mk}}{\partial x_{l}'}(x,x';t-t') \: .
\label{iden}
\end{equation}
A single dislocation loop is considered to be 
present. At the position of the loop itself
the derivatives of the displacement field have singularities, and 
across the slip plane the
displacement field has a discontinuity equal to the Burgers vector
$b_i$.
The identity in Eq. (\ref{iden}) is integrated over the volume of the 
elastic sample,
excluding a thin tube around the dislocation loop, as well as a thin
layer encompassing the slip plane. Using the reciprocity relation of the
impulse response (Poruchikov, 1993):
\begin{equation}
G_{im}(x,x';t-t') = G_{mi}(x',x;t-t') \: ,
\end{equation}
as well as Eqns. (\ref{eq:wave})-(\ref{eq:bcg}), and integrating by
parts, the following identity results:
\begin{eqnarray}
\int dS_{j}'C_{ijkl}\frac{\partial U_{k}}{\partial x_{l}'}(x',t')
G_{mi}(x,x';t-t') - \int dx' C_{ijkl}\frac{\partial^{2} U_{k}}{
\partial x_{l}' \partial x_{j}'}(x',t')G_{mi}(x,x';t-t') = \nonumber \\
\int dS_{j}' C_{ijkl}U_{i}(x',t')\frac{\partial G_{mk}}{\partial x_{l}'}
(x,x';t-t') - \int dx' C_{ijkl}U_{i}(x',t')\frac{\partial^{2}G_{mk}}
{\partial x_{l}'\partial x_{j}'}(x,x';t-t')
\: , \label{ini}
\end{eqnarray}
where the surface integrals are carried out over the tube and layer just 
introduced, as well as over the sample boundary.
The first term on the left vanishes on the surfaces
of the sample due to the free surface boundary condition, Eq. (\ref{bc}),
and it also vanishes on the slip plane due to
continuity of the stress there.  The first term on the right also vanishes
on the surfaces of the sample due to the free surface boundary  
condition, Eq. (\ref{eq:bcg}), due to reciprocity,  but it will give
a contribution on the slip plane due to the discontinuity of $U_{i}$
 given by the Burgers vector $b_{i}$.
The first terms on the left and on the right
of Eq. (\ref{ini}) integrated over the small tube (of radius $\epsilon$)
around the dislocation loop vanish because of the short distance  
behaviour of particle displacement $U_i$ (Lund, 1988). 

Using
the differential equations satisfied by $U_{i}$ and $G_{mi}$
one gets from Eq. (\ref{ini}):
\begin{eqnarray}
-\rho \int dx' \frac{\partial^{2}U_{i}(x',t')}{\partial t^{'2}}
G_{mi}(x,x';t-t') = b_{i} \int_{S(t')} dS_{j}' C_{ijkl}
\frac{\partial G_{mk}}{\partial x_{l}'}(x,x';t-t') \nonumber \\
-\rho \int dx' \frac{\partial^{2} G_{mi}}{\partial t^{'2}}
(x,x';t-t')U_{i}(x',t') + \delta (t-t') U_{m}(x,t') \: ,
\end{eqnarray}
with $S(t')$ the slip plane at time $t'$.
Integrating over time $t'$, one obtains the expression for the displacement
as an integral over the succesive slip planes:
\begin{equation}
U_{m}(x,t) = -b_{i} \int_{-\infty}^{\infty} dt' \int_{S(t')}
dS_{j}' C_{ijkl} \frac{\partial G_{mk}}{\partial x_{l}'}(x,x';t-t')
\: .
\label{ummp}
\end{equation}
Using reciprocity, this can be written as
\beq
U_{m}(x,t) = -b_{i} \int_{-\infty}^{\infty} dt' \int_{S(t')}
dS_{j}' \sigma_{ij}^{Gm}(x',x;t-t') \: .
\label{umm}
\eeq
Attention should be paid to the order in which the variables appear
in this formula.
In an analogous way, for the stationary case, the following 
displacement field is derived:   
\beq
U_{m}(x) = -b_{i}  \int_{S'}
dS_{j}' \sigma_{ij}^{Gm}(x',x) \: ,
\label{ums}
\eeq
with the static Green's function satisfying:
\beq
C_{ijkl}\frac{\partial^{2}}{\partial x_{k}
\partial x_{j}} G_{lm}(x,x') =
\frac{\partial}{\partial x_{j}} \sigma_{ij}^{Gm}(x,x') =
-\delta_{im}\delta(x-x') \: ,
\label{egs}
\eeq
and the free surface boundary condition:
\beq
\sigma_{ij}^{Gm}(x_{S},x')n_{j}(x_{S})=0 \: .
\label{bcs}
\eeq
Eq. (\ref{ummp}) leads to null normal stresses at the free surfaces.
Indeed, from Eq. (\ref{ummp}) one gets
\bea
\sigma_{pq}(x,t) & = & -b_{i}C_{pqnm} \int_{-\infty}^{\infty}
dt' \int_{S(t')} dS_{j}' C_{ijkl}
\frac{\partial^{2} G_{mk}}{\partial x_{n} \partial x_{l}'}
(x,x';t-t') \nonumber \\
& = & -b_{i}C_{ijkl} \int_{-\infty}^{\infty}
dt' \int_{S(t')} dS_{j}'
\frac{\partial \sigma_{pq}^{Gk}}{\partial x_{l}'}(x,x';t-t')
\: .
\eea
Since $\sigma_{pq}^{Gk}(x_{S},x';t-t')n_{q}(x_{S})=0$, this implies
that $\sigma_{pq}(x_{S},t)n_{q}(x_{S})=0$, as it should.

\section{Stress fields due to dislocation loops in finite samples}

\subsection{Stationary case}
Starting from Eq. (\ref{ums}), the
stress field produced by the static dislocation loop is:
\beq
\sigma_{pq}(x) = -b_{i}C_{pqlm} \int_{S'} dS_{j}' 
\frac{\partial}{\partial x_{l}} 
 \sigma_{ij}^{Gm}(x',x) \: .
\label{sip}
\eeq
In the infinite medium case, space homogeneity easily leads to a line 
integral representation (Mura, 1963).
In order to obtain a line integral representation for this stress
field in the finite case, we add and substract to the integrand the term 
\[ 
\frac{\partial}{\partial x_{l}'} 
 \sigma_{ij}^{Gm}(x',x) , 
 \]
which is easy to turn into a line integral, and we define a tensor:
\beq
f_{ij}^{lm}(x',x) \equiv \frac{\partial}{\partial x_{l}}
\sigma_{ij}^{Gm}(x',x)+\frac{\partial}{\partial x_{l}'}
\sigma_{ij}^{Gm}(x',x) \: , 
\label{eq:fdef}
\eeq
which is divergenceless. Indeed, from Eq. (\ref{egs}),
\beq
\frac{\partial }{\partial x_{j}'} f_{ij}^{lm}(x',x) =
- \delta_{im} (\frac{\partial}{\partial x_{l}} +
\frac{\partial}{\partial x_{l}'}) \delta (x-x') =0 \: .
\eeq
This means that $f_{ij}^{lm}(x',x)$ can be written as the curl of a
``vector potential'' $A_{s}^{ilm}(x',x)$:
\beq
f_{ij}^{lm}(x',x) =  \epsilon_{jrs} \frac{\partial}{\partial
x_{r}'} A_{s}^{ilm}(x',x) \: ,
\label{curl}
\eeq
and it is possible to write the stress field, Eq. (\ref{sip}), as:
\beq
\sigma_{pq}(x) = -b_{i} C_{pqlm} \int_{S'} dS_{j}'
\{ f_{ij}^{lm}(x',x)-\frac{\partial \sigma_{ij}^{Gm}}{
\partial x_{l}'}(x',x) \} \: .
\eeq
When $x$ is not in the slip plane $S'$
the following is true:
\beq
\int dS_{j}' \frac{\partial \sigma_{ij}^{Gm}}{\partial x_{l}'}
(x',x) = \epsilon_{vjl}\epsilon_{vsr} \int dS_{s}' \frac{\partial}
{\partial x_{r}'} \sigma_{ij}^{Gm}(x',x) \: ,
\eeq
so that we finally have that the stress field 
can be expressed as a line integral along the dislocation loop:
\beq
\sigma_{pq}(x) = b_{i} C_{pqlm} \int_{L'} dl_{s}'
\{ \epsilon_{sjl} \sigma_{ij}^{Gm}(x',x) -
A_{s}^{ilm}(x',x) \} \: ,
\label{fin}
\eeq
as promised,
with $A_{s}^{ilm}(x',x)$ a ``vector potential'' to be determined 
for each specific sample by use of  Eqns. (\ref{eq:fdef}) and  
(\ref{curl}). This means that $\sigma_{ij}^{Gm}$, the Green's function  
for the finite sample, must be known. Its determination, however, is of  
course independent of the particular dislocation loop that might be  
under consideration.

Eqn. (\ref{fin}) is similar to Eqn. (12) of Gosling and Willis (1994). The 
difference between those two expressions resides in the fact that Gosling 
and Willis (1994) split the integrand into the Green's function for an 
infinite medium plus ``image'' forces that annul the tractions at the free 
surface.

\subsection{Non stationary case}

Since the problem is
homogeneous under time translations it is possible to obtain an
expression for particle velocity, 
that involves an integral along the dislocation loop
only.
Indeed, 
from Eq. (\ref{umm}):
\beq
\frac{\partial U_{m}}{\partial t}(x,t)  = 
b_{i} \int_{-\infty}^{\infty} dt' \{ \frac{\partial}{\partial t'} [
\int_{S(t')} dS_{j}'  \sigma_{ij}^{Gm}
(x',x;t-t') ] 
 - \int_{\frac{dS}{dt'}(t')}
dS_{j}' \sigma_{ij}^{Gm}(x',x;t-t') \}
\: .
\eeq
The first term is zero since the Green's function vanishes at
$t-t'=\pm \infty$.
Since: 
\[
\int_{dS(t')/dt'}dS_{j}' = \epsilon_{jpq} \int_{L(t')} dl_{q}'
V_{p}(x',t') \: ,
\]
where $L(t')$ is the dislocation loop bounding the slip
plane $S(t')$, $V_p(x',t')$ is the loop's local
 velocity and $\epsilon_{jpq}$ the completely
antisymmetric tensor in three dimensions; we then have:
\begin{equation}
\frac{\partial U_{m}}{\partial t}(x,t) = -b_{i} 
\int_{-\infty}^{\infty} dt'
\int_{L(t')} dl_{q}' \sigma_{ij}^{Gm}(x',x;t-t') \epsilon_{jpq}
V_{p}(x',t') \: .
\end{equation}
From this line integral expression 
for the displacement velocity field, one 
obtains an expression for the partial time derivative of the 
stress field that involves line integrals over the dislocation loop:
\beq
\frac{\partial}{\partial t}
\sigma_{ln}(x,t) = -C_{lnkm}b_{i} \int_{-\infty}^{\infty} dt'
\int_{L(t')} dl_{q}' \frac{\partial}{\partial x_{k}} 
\sigma_{ij}^{Gm}(x',x;t-t') \epsilon_{jpq}
V_{p}(x',t') \: .
\label{pasi}
\eeq
This result, together with the
result for a static loop (considered as an initial condition),
mean that the time dependent stress field can be written in terms
of integration of line integrals over the dislocation loop, 
and it is independent of the choice of slip planes, as it should.
Explicitly, we are thinking on the generally valid formula:
\beq
\sigma_{ln}(x,t) = \sigma_{ln}(x,t=0) + \int_{0}^{t} dt'
\frac{\partial \sigma_{ln}}{\partial t} (x,t') \: ,
\eeq
with both terms on the right hand side having a line integral 
representation.

A general result that derives from Eq. (\ref{pasi}) is that segments
of the dislocation loop on the free surfaces do not contribute to
$\partial \sigma_{ln}/\partial t$: in those segments $d\vec{l}' \times
\vec{V}$ is parallel to the normal $\hat{n}$
to the free surface, and then the boundary
condition satisfied by $\sigma_{ij}^{Gm}$ 
at the free surfaces cancels that contribution.

\section{Examples of explicit ``Vector potentials'' \\ for the stationary 
case}
 
In two geometries, a half space and a thin plate, and considering
an isotropic elastic medium we have determined
 explicitly  the form of the ``vector potentials'' appearing in
the general line integral representation of Eq. (\ref{fin}) for the 
stress field generated by a dislocation in a finite sample. The
derivation will be explained in the following, with further details
to be found in Appendices I and II.

The elastic medium fills the half space  $z \geq 0$, and the thin
plate the space between $-h \leq z \leq h$, where $z$ is the coordinate 
whose
direction is perpendicular to the free surfaces of these geometries
(we will call $\vec{R}$ a coordinate in the planes parallel to 
the free surfaces).
In both geometries the appropriate Green's function depends on 
$\vec{R}-\vec{R}'$ because of homogeneity in those directions, meaning
that only differentiation with respect to $z$ and $z'$ can give
a non zero result for the ``vector potentials'' of Eq. (\ref{curl});
in other words, only the component $l=z$ of $A_{s}^{ilm}$ is non zero. 
The equation to be solved for $A_{s}^{izm}$ is then:
\beq
\frac{\partial}{\partial z}
\sigma_{ij}^{Gm}(z',z;\vec{R}'-
\vec{R})+\frac{\partial}{\partial z'}
\sigma_{ij}^{Gm}(z',z;\vec{R}'-\vec{R}) =
\epsilon_{jrs} \frac{\partial}{\partial
x_{r}'} A_{s}^{izm}(z',z;\vec{R}'-\vec{R}) \: .
\label{eqa}
\eeq
We have obtained the Green's function appropriate for  
these geometries, satisfying Eqs.
(\ref{egs}) and (\ref{bcs}), using the method
of Chishko (1989), who found the time dependent 
Green's function for a thin plate. The Green's functions are written as
the infinite medium Green's function ($G_{im}^{(0)}$)
minus a term ($H_{im}$) that satisfies
the homogeneous equilibrium equation and that cancels the normal
stresses produced by the infinite medium's Green's function at the 
free surfaces:
\beq
G_{im}(z,z';\vec{R}-\vec{R}')=G_{im}^{(0)}(x-x')-H_{im}(z,z';
\vec{R}-\vec{R}') \: .
\label{gims}
\eeq
This strategy differs somewhat from the one emplyed by Gosling and Willis 
(1994), who used a similar split not for the Green's function, but for a 
combination of gradients of the Green's function.

The representation used for $H_{im}$ for the half space is the following:
\beq
H_{im}(z,z';\vec{R}-\vec{R}') = -\int d\vec{R}'' G_{in}^{(h)}
(z;\vec{R}-\vec{R}'') \sigma_{nz}^{(0)m}(-z';\vec{R}''-\vec{R}') \: ,
\label{hims}
\eeq
and a similar form for the thin plate:
\beq
H_{im}(z,z';\vec{R}-\vec{R}') = \sum_{a=1,2} \int dS_{a}'' G_{in}^{(p)}
(z',\xi^{(a)};\vec{R}'-\vec{R}'') \sigma_{nz}^{(0)m}(
\xi^{(a)}-z;\vec{R}''-\vec{R}) \: ,
\label{himp}
\eeq
where the index $(a)=1,2$ represents the free surfaces at $z''=
\xi^{(a)}=\pm h$, and
$dS_{a}''= \pm d\vec{R}''$.
We have introduced the surface Green's functions $G_{in}^{(h),(p)}$ for
the half space and the thin plate: they are the displacements
in direction $(i)$ produced by a localized impulse in direction $(n)$
applied at the free surface. They satisfy, respectively,  
the homogeneous equilibrium equations:
\bea
C_{ijpq}\frac{\partial^{2}}{\partial x_{p} \partial x_{j}}
G_{qn}^{(h)}(z;\vec{R}-\vec{R}') & = & 0 \label{egh} \\
C_{ijpq}\frac{\partial^{2}}{\partial x_{p} \partial x_{j}}
G_{qn}^{(p)}(z,\xi;\vec{R}-\vec{R}') & = & 0  
\label{egp}
\eea
($\xi=\pm h$), 
and they are subject to the boundary conditions:
\bea
\sigma_{iz}^{(h)n}(z=0;\vec{R}-\vec{R}') & = & - \delta_{in} \delta(\vec{R}-
\vec{R}') \label{bgh} \\
\sigma_{iz}^{(p)n}(\xi^{(b)},\xi^{(a)};\vec{R}-\vec{R}') & = & 
\pm \delta^{(a)(b)}\delta_{in} \delta(\vec{R}-
\vec{R}') \label{bgp} \: .
\eea 

It is possible to solve for the 
Fourier components (in the plane parallel to
the free surface) of the Green's functions $G_{ln}^{(h)}(z;\vec{R}-
\vec{R}')$
($g_{ln}^{(h)}(z|\vec{k})$),
and $G_{ln}^{(p)}(z,\xi;\vec{R}-
\vec{R}')$
($g_{ln}^{(p)}(z,\xi|\vec{k})$): 
\bea 
g_{ln}^{(h)}(z|\vec{k}) & = & \int d\vec{R}
e^{i \vec{k} \cdot \vec{R}} G_{ln}^{(h)}(z;\vec{R}) \nonumber \\
g_{ln}^{(p)}(z|\vec{k}) & = & \int d\vec{R}
e^{i \vec{k} \cdot \vec{R}} G_{ln}^{(p)}(z;\vec{R}) \: ,
\eea
and they are given in Appendices I and II. From Eqs. 
(\ref{hims}) and (\ref{himp}), one obtains in Fourier space:
\bea
\sigma_{ij}^{H(h)m}(z',z|\vec{k}) & = & - \sigma_{ij}^{(h)n}
(z'|\vec{k})\sigma_{nz}^{(0)m}(-z|\vec{k}) \label{shh} \\
\sigma_{ij}^{H(p)m}(z',z|\vec{k}) & = & \sum_{\xi} (\pm) \sigma_{ij}^{(p)n}
(z',\xi|\vec{k})\sigma_{nz}^{(0)m}(\xi-z|\vec{k}) \: ,
\label{shp}
\eea
with $\xi = \pm h$.
 Expressions for $\sigma_{ij}^{H(h)m}
(z'|\vec{k})$ and $\sigma_{nz}^{(0)m}(z-z'|\vec{k})$ are shown
also in Appendix I, and for $\sigma_{ij}^{H(p)m}
(z',\xi|\vec{k})$ in Appendix II.

Since $\sigma_{ij}^{(0)m}$ depends on $(z'-z)$, only the
term $\sigma_{ij}^{Hm}$ contributes to the determination of the
``vector potentials'' in Eq. (\ref{eqa}). If $a_{s}^{im}(z',z;\vec{R}'-
\vec{R})$ is the ``vector potential'' associated with 
$\sigma_{ij}^{Hm}$, namely
\beq
\sigma_{ij}^{Hm}(z',z;\vec{R}'-\vec{R}) = \epsilon_{jrs}\frac{
\partial a_{s}^{im}}{\partial x_{r}'}(z',z;\vec{R}'-\vec{R}) \: ,
\eeq
then the ``vector potential'' $A_{s}^{izm}$ that satisfies Eq. (\ref{eqa}) 
is
given by:
\beq
A_{s}^{izm}(z',z;\vec{R}'-\vec{R}) = -(\frac{\partial}{\partial z}+
\frac{\partial}{\partial z'})a_{s}^{im}(z',z;\vec{R}'-\vec{R}) \: .
\label{aa}
\eeq
For the sake of brevity, we will present the stress function tensor 
associated with $\sigma_{ij}^{Hm}$, from which $a_{s}^{im}$ and 
further $A_{s}^{izm}$ can be determined by differentiation. The
stress function tensor $\chi_{qn}^{m}$ is defined through a ``double
curl'':
\beq
\sigma_{ij}^{Hm}(z',z;\vec{R}'-\vec{R}) = -\epsilon_{ipq}
\epsilon_{jln}\frac{\partial^{2} \chi_{qn}^{m}}{
\partial x_{p}' \partial x_{l}'}(z',z;\vec{R}'-\vec{R}) \: ,
\eeq
a form that assures that $\sigma_{ij}^{Hm}$, symmetric in $(ij)$, is 
divergenceless in the indices $i$ and $j$, as it should. 
It is easy to see that:
\beq
a_{s}^{im}(z',z;\vec{R}'-\vec{R}) = -\epsilon_{ipq} 
\frac{\partial \chi_{qs}^{m}}{\partial x_{p}'}(z',z; \vec{R}'-
\vec{R}) \: .
\label{adef}
\eeq
In Fourier space the stress function tensor for the 
isotropic half space 
geometry is:
\bea
\chi_{\eta \nu}^{z}(z',z|\vec{k}) & = & 
-\frac{e^{-k(z+z')}}{2 \gamma^{2}(\gamma^{2}-1)k^{4}}
\{ (\gamma^{2}-2)[ (\gamma^{2}+1) +
2(\gamma^{2}-1)kz ] k_{\eta}k_{\nu} \nonumber \\ & & - 
(\gamma^{2}-1)[\gamma^{2}+(\gamma^{2}-1)k z +
(\gamma^{2}+1) kz' +2(\gamma^{2}-1)k^{2}zz' ] q_{\eta}q_{\nu} \}
\nonumber \\
\chi_{\eta \nu}^{\beta}(z',z|\vec{k}) & = & 
\frac{i e^{-k(z+z')}}{2 \gamma^{2}(\gamma^{2}-1)k^{5}}
\{ \gamma^{2}(\gamma^{2}-1)q_{\beta}(q_{\eta}k_{\nu}+
q_{\nu}k_{\eta}) \nonumber \\ & & 
+ (\gamma^{2}-1)[1-(\gamma^{2}-1)k z +
(\gamma^{2}+1) kz' -2(\gamma^{2}-1)k^{2}zz' ]k_{\beta} q_{\eta}q_{\nu}
\nonumber \\ & &
- (\gamma^{2}-2)[ (\gamma^{2}+1) -
2(\gamma^{2}-1)kz ] k_{\beta}k_{\eta}k_{\nu}
 \} \: , 
\label{sfth}
\eea
where $q_{\nu} \equiv \epsilon_{\nu \delta}
k_{\delta}$, greek indices mean coordinates $x$ or $y$, and $\gamma^{2} 
\equiv
c_{l}^{2}/c_{t}^{2}$ is the ratio between the longitudinal and transverse
sound velocities in an isotropic medium. 
These formulae were obtained using the explicit forms for the
Fourier components $\sigma_{ij}^{Hm}(z',z|\vec{k})$ that are
written in Appendix I,  Eq. (\ref{shf}),  and solving for a stress
function tensor that has only components $\chi_{\eta \nu}^{m}
\neq 0$ in order to reproduce $\sigma_{ij}^{Hm}$. Similarly for a plate:
\bea
\chi_{\eta \nu}^{z}(z',z|\vec{k}) & = & \sum_{\xi} (\pm)
\frac{e^{-k|\xi-z|}}{4 \gamma^{2}(\gamma^{2}-1)k^{4}}
\{ (\gamma^{2}-1)[(1+(\gamma^{2}-1)k|\xi -z|)(
kz' \phi_{4}-k\xi \phi_{3}) 
\nonumber \\ & & - s(\xi -z) (\gamma^{2}+
(\gamma^{2}-1)k|\xi-z|)(\phi_{3}+
k\xi \phi_{1}-kz'\phi_{2}) ] q_{\eta}q_{\nu}
\nonumber \\ &  & + (\gamma^{2}-2)[ s(\xi-z)(
\gamma^{2}+(\gamma^{2}-1)k|\xi-z|) \phi_{3}+
(1+(\gamma^{2}-1)k|\xi-z|)\phi_{1}] k_{\eta}k_{\nu} \} \nonumber \\
\chi_{\eta \nu}^{\beta}(z',z|\vec{k}) & = & \sum_{\xi} (\pm) 
 \frac{i e^{-k|\xi-z|}}{4 \gamma^{2}(\gamma^{2}-1)k^{5}}
\{ (\gamma^{2}-1)[(1-(\gamma^{2}-1)k|\xi -z|)(\phi_{3}
+k \xi \phi_{1}-kz' \phi_{2}) 
\nonumber \\ & & - s(\xi -z) (\gamma^{2}-
(\gamma^{2}-1)k|\xi-z|)(
kz' \phi_{4}-k\xi \phi_{3}) ] k_{\beta}q_{\eta}q_{\nu}
\nonumber \\ &  & - (\gamma^{2}-2)[ s(\xi-z)(
\gamma^{2}-(\gamma^{2}-1)k|\xi-z|) \phi_{1}+
(1-(\gamma^{2}-1)k|\xi-z|)\phi_{3}] k_{\beta}k_{\eta}k_{\nu} 
\nonumber \\
&  & + s(\xi-z) \gamma^{2}(\gamma^{2}-1)( \frac{\cosh(kz')}
{\sinh(k\xi)}+\frac{\sinh(kz')}{\cosh(k\xi)})q_{\beta}
(q_{\eta}k_{\nu}+q_{\nu}k_{\eta}) \} \: .
\label{sftp}
\eea

\section{Example: a screw dislocation perpendicular to 
the free surfaces of a half space and a thin plate}

Eshelby and Stroh (1951)  obtained the elastic fields
corresponding to a stationary screw dislocation perpendicular to
the free surfaces of a thin plate. From their method it is easy to
obtain the corresponding fields for a half space. As examples of
the use of the general line integral representation for the
stresses presented in this paper,
Eq. (\ref{fin}), we rederive these exact 
results. In the following we go through the main steps in the 
calculation.

\subsection{A screw dislocation in a half space}

Using the method in Eshelby and Stroh (1951), a simple calculation gives
 the displacement and stress fields, in polar
coordinates, due to a screw dislocation lying along the $z$ axis 
perpendicular
to the free surface of a half space $z > 0$:
\begin{eqnarray}
U_{z}(\theta) & = & \frac{b}{2 \pi} \theta \nonumber \\
U_{\theta}(R,z) & = & \frac{b}{2 \pi} \int_{0}^{\infty} \frac{dk}{k}
e^{-k z} J_{1}(kR) \nonumber \\
\sigma_{z \theta}(R,z) & = & \frac{\mu b}{2 \pi R} -
\frac{\mu b}{2 \pi} \int_{0}^{\infty} dk e^{-k z}
J_{1}(kR) \nonumber \\
\sigma_{r \theta}(R,z) & = & -\frac{\mu b}{2 \pi}
\int_{0}^{\infty} dk e^{-k z} J_{2}(kR) \: ,
\label{sigh}
\end{eqnarray}
with $J_{1}(z)$ and $J_{2}(z)$ the Bessel functions of order one
and two respectively, and $\mu$ the shear modulus.

Now we describe how these results can be obtained as a special case of
the general formula, Eq. (\ref{fin}). When applied to this case it reads: 
\beq
\sigma_{pq}(x) = \sigma_{pq}^{(1)}(x) +\sigma_{pq}^{(2)}(x) \: ,
\eeq
where:
\bea
\sigma_{pq}^{(1)}(z,\vec{R}) & \equiv & 
b C_{pq\gamma m}\epsilon_{\eta \gamma }\int_{0}^{\infty} dz' 
 \sigma_{z \eta}^{Gm}(z',z;-\vec{R})
 \label{sze1} \\ \sigma_{pq}^{(2)}(z,\vec{R}) & \equiv &
-bC_{pqmz} \int_{-C} dR_{\beta}' A_{\beta}^{zzm}(z'=0,z;\vec{R}'
-\vec{R}) \: ,
\label{sze2}
\eea
with $-C$ a curve that closes the dislocation loop on the free surface,
and $\epsilon_{12}=-\epsilon_{21}=1$, $\epsilon_{11}=\epsilon_{22}=0$.

First we focus on the first term $\sigma_{pq}^{(1)}(x)$. For this term
the segment on the free surface doesn't contribute due to the boundary
condition: $\sigma_{zi}^{Gm}(x_{S}',x)=0$. In Fourier space  
Eq. (\ref{sze1}) becomes:
\beq
\sigma_{pq}^{(1)}(z,-\vec{k}) = b C_{pq \gamma m} 
\epsilon_{\eta \gamma} \int_{0}^{\infty}
dz' \sigma_{z \eta}^{Gm}(z',z|\vec{k}) \: .
\eeq
After doing the integration with the expressions in Appendix I,
one obtains $\sigma_{zz}^{(1)}(z|\vec{k})=0$, and:
\beq
\sigma_{z \alpha}^{(1)}(z|\vec{k}) = -ib\mu \frac{\epsilon_{\alpha \eta} 
k_{\eta}}{k^{2}}  \: . 
\eeq
Going back to real space, and in polar coordinates, we get
\beq
\sigma_{z \theta}^{(1)}(z|\vec{R}) = \frac{ib\mu}{(2\pi)^{2}}
\int d\vec{k} \frac{e^{-i \vec{k} \cdot \vec{R}}}{k}
(\sin \theta \sin \alpha +\cos \theta \cos \alpha ) \: ,
\eeq
where $\alpha$ is the polar angle associated with $\vec{k}$, 
and $\theta$ the angle of $\vec{R}$ with respect to some fixed axis. 
Defining $\phi=\alpha-\theta$, one gets effectively an integral over
$\phi$, which is:
\beq
\sigma_{z \theta}^{(1)}(z|\vec{R}) = 
\frac{ib \mu}{(2 \pi)^{2}} \int_{0}^{\infty} dk
\int_{-\pi}^{\pi} d\phi e^{-i kR \cos \phi} \cos \phi = \frac{b \mu}{2\pi} 
 \int_{0}^{\infty} dk J_{1}(kR) = \frac{\mu b}{2 \pi R} \: ,
\eeq
i.e. it reproduces the first exact term of $\sigma_{z \theta}(R,z)$ in Eq.
(\ref{sigh}). Similarly it can be shown that $\sigma_{z r}^{(1)}(R,z)=0$.
Also, this type of term is the only contributing to the calculation
of $\sigma_{r \theta}(R,z)$, and by a similar integration one gets the
exact result in Eq. (\ref{sigh}).

Now we focus on the second term $\sigma_{pq}^{(2)}(x)$. For this term,
only the segment on the surface contributes. 
From expressions in Appendix I, we see that 
the ``vector potential'' $A_{\beta}^{z z z}$ is zero at the free surface,
and that $A_{\beta}^{z z \delta}$ is given 
in Fourier space by:
\beq
A_{\beta}^{z z \delta}(z'=0,z;\vec{k})=
\frac{e^{-k z}}{k^{2}}k_{\beta}\epsilon_{\delta \eta} k_{\eta}
\: .
\eeq
Now, the integral $\int_{-C} dR_{\beta}' A_{\beta}^{zz \delta}(x',x)$ 
appearing in Eq. (\ref{sze2}) is
independent of the choice of curve $-C$ on the free surface because the
component $z$ of the curl of $A_{j}^{zz \delta}$ with respect to
the index $j$ is zero on the free surface: indeed, $ik_{\gamma}
\epsilon_{\gamma \beta}A_{\beta}^{zz\delta}(z'=0,z|\vec{k})=0$.
Thus, the expression for $\sigma_{z \delta}^{(2)}(x)$ in Eq. (\ref{sze2})
becomes:
\beq
\sigma_{z \delta}^{(2)}(x) = -\frac{b \mu}{(2 \pi)^{2}} 
\int_{-\pi}^{\pi} d\alpha \int_{-C} d\vec{R}' \cdot \vec{k} 
\int_{0}^{\infty} \frac{dk}{k} e^{-i k
|\vec{R}'-\vec{R}| \cos \phi} \epsilon_{\delta \eta}
k_{\eta} e^{-kz} \: ,
\eeq
with $\alpha$ the angle of $\vec{k}$ and $\phi$ the angle between
$\vec{k}$ and $(\vec{R}'-\vec{R})$.
The curve $C$ was chosen in a convenient way. First, it starts at 
$\vec{R}'=0$,
which coincides with the position of the ``vertical'' segment of the 
dislocation. Then it continues in a semicircle around the point $\vec{R}$
(there $|\vec{R}'-\vec{R}|=R$).
After that, it continues on a straight line parallel to the direction of
the vector $\vec{R}$ ( $\vec{R}'-\vec{R}=\lambda \vec{R}$, $\lambda \in
(1,\infty)$).
In this way, the result from both of these segments
adds up to the result:
\beq
\sigma_{z \theta}^{(2)}(R,z) = -\frac{b \mu}{2 \pi} \int_{0}^{\infty}
dk e^{-k z} J_{1}(kR) \: ,
\eeq
which is the second term of the exact result in Eq. (\ref{sigh}), as it
remained to be proven.

\subsection{A screw dislocation in a plate} 
Eshelby and Stroh (1951) obtained 
the exact displacement and stress fields, in polar
coordinates, due to a screw dislocation along the $z$ axis perpendicular
to the free surfaces of a plate (the
plate exists for $-h < z < h$):
\begin{eqnarray}
U_{z}(\theta) & = & \frac{b}{2 \pi} \theta \nonumber \\
U_{\theta}(R,z) & = & -\frac{b}{2 \pi} \int_{0}^{\infty} \frac{dk}{k}
\frac{\sinh (kz)}{\cosh (kh)} J_{1}(kR) \nonumber \\
\sigma_{z \theta}(R,z) & = & \frac{\mu b}{2 \pi R} -
\frac{\mu b}{2 \pi} \int_{0}^{\infty} dk \frac{\cosh (kz)}{\cosh (kh)}
J_{1}(kR) \nonumber \\
\sigma_{r \theta}(R,z) & = & \frac{\mu b}{2 \pi}
\int_{0}^{\infty} dk \frac{\sinh (kz)}{\cosh (kh)} J_{2}(kR) \: .
\label{sigp}
\end{eqnarray}
The verification that these results follow as a special case of 
Eq. (\ref{fin}) is analogous to the half-space case:
one needs the ``vector potential'' $A_{\beta}^{z z \delta}$ evaluated at
the free surfaces, which in Fourier space is, from Appendix I,
\beq
A_{\beta}^{z z \delta}(\pm h,z;\vec{k})  = 
-\frac{k_{\beta}\epsilon_{\delta \nu} k_{\nu}}{2k^{2}}
\{ \pm \frac{\cosh(kz)}{\cosh(kh)}  +
\frac{\sinh(kz)}{\sinh(kh)} \} \: .
\eeq
The terms $\sigma_{pq}^{(1)}(x)$ and $\sigma_{pq}^{(2)}(x)$
from Eq. (\ref{fin}) become:
\bea
\sigma_{pq}^{(1)}(z,\vec{R}) & \equiv &
b C_{pq\gamma m}\epsilon_{\eta \gamma }\int_{-h}^{h} dz'
 \sigma_{z \eta}^{Gm}(z',z;-\vec{R})
 \label{szp1} \\ 
\sigma_{pq}^{(2)}(z,\vec{R}) & \equiv &
2bC_{pqmz} 
\frac{1}{(2\pi)^{2}} \int d\vec{k} 
\int_{C} dR_{\beta}' k_{\beta}
e^{-i \vec{k} \cdot (
\vec{R}'-\vec{R}) }
\frac{\epsilon_{\delta \nu} k_{\nu}
\cosh (kz)}{2k^{2} \cosh (kh) } \: .
\label{szp2}
\eea
The necessary integrations are done similarly as for the half space
case (the curve $C$ is the same), and the only non zero components
of the stress that are obtained are those given by Eq. (\ref{sigp}).
($\sigma_{pq}^{(1)}(x)$ contributes the first term of $\sigma_{z \theta}$,
and $\sigma_{pq}^{(2)}(x)$ the second term of $\sigma_{z \theta}$).

\section{Discussion}
We have given a line integral representation for the stresses of an 
arbitrary dislocation loop in an arbitrary, not necessarily isotropic, three 
dimensional, finite elastic body. This representation is given in terms of a 
``vector potential'', for whose computation the Green's function for the 
elastic body in question must be known. This, in general, is not the case, 
but use of approximate expressions will yield approximations to the 
corresponding stresses. At any rate, we have provided explicit forms of 
the said 
``vector potentials'' for a half space and for a thin plate. Previous 
results for the stresses due to a screw dislocation in a thin plate (Eshelby 
and Stroh, 1951), as well as easily obtained stresses for a screw 
dislocation perpendicular to the surface of a half space, were recovered as 
special cases.

Gosling and Willis (1994) have provided an alternative line integral 
representation for the stress due to an arbitrary dislocation in an 
isotropic half-space. Our formulation is more general, admittedly at the 
price of introducing the Green's function for a finite elastic sample, a 
quantity that is in general hard to compute.

\section*{Acknowledgments}

This work was supported in part by the Andes Foundation, Fondecyt
Grants 3950011 and 196082, and a C\'atedra Presidencial en Ciencias.

\section*{Appendix I: Green's function of a half space}
We write the Fourier components of the surface Green's function appropriate
for a half space, $g_{ij}^{(h)}(z|\vec{k})$ (they are the solutions in 
Fourier space of Eq. (\ref{egh}) subject to the boundary condition
given by Eq. (\ref{bgh})) :
\bea
g_{zz}^{(h)}(z|\vec{k}) & = & \frac{e^{-k z}}{2 \mu (\gamma^{2}-1)k}
\{\gamma^{2}+(\gamma^{2}-1)k z \} \nonumber \\
g_{z \beta}^{(h)}(z|\vec{k}) & = & \frac{i k_{\beta}e^{-k z}}
{2 \mu (\gamma^{2}-1)k^{2}}
\{1+(\gamma^{2}-1)k z \} \nonumber \\
g_{\beta z}^{(h)}(z|\vec{k}) & = & \frac{i k_{\beta}e^{-k z}}
{2 \mu (\gamma^{2}-1)k^{2}}
\{-1+(\gamma^{2}-1)k z \} \nonumber \\
g_{\alpha \beta}^{(h)}(z|\vec{k}) & = & \frac{e^{-k z}}
{2 \mu (\gamma^{2}-1)k^{3}}
\{2 (\gamma^{2}-1) q_{\alpha}q_{\beta}+
[\gamma^{2}-(\gamma^{2}-1)k z]k_{\alpha}k_{\beta} \} \: ,
\eea
where $q_{\alpha} \equiv \epsilon_{\alpha \nu} k_{\nu}$ is a vector
perpendicular to $k_{\nu}$ ($\epsilon_{12}=1$, $\epsilon_{21}=-1$).
The Fourier components of $\sigma_{iz}^{(0)m}(z-z'|\vec{k})$ are:
\bea
\sigma_{zz}^{(0)z}(z-z'|\vec{k}) & = & -\frac{s(z-z')}
{2 \gamma^{2}}e^{-k|z-z'|}
\{ \gamma^{2} +(\gamma^{2}-1)k|z-z'| \} \nonumber \\
\sigma_{\beta z}^{(0)z}(z-z'|\vec{k}) & = & -\frac{i k_{\beta}}
{2 \gamma^{2}k}e^{-k|z-z'|} \{ 1+ (\gamma^{2}-1)k |z-z'| \} \nonumber \\
\sigma_{z z}^{(0)\delta}(z-z'|\vec{k}) & = & \frac{i k_{\delta}}
{2 \gamma^{2}k}e^{-k|z-z'|} \{ 1- (\gamma^{2}-1)k |z-z'| \} \nonumber \\
\sigma_{\beta z}^{(0)\delta}(z-z'|\vec{k}) & = & -\frac{s(z-z')}
{2 \gamma^{2} k^{2}}e^{-k|z-z'|} \{ \gamma^{2} q_{\beta}q_{\delta} +
[\gamma^{2}-(\gamma^{2}-1)k |z-z'|]
k_{\beta}k_{\delta} \} \: . \nonumber \\
\eea
Doing the calculation in Eq. (\ref{shh}), one gets the Fourier components
$\sigma_{ij}^{H m}(z',z|\vec{k})$:
\bea
\sigma_{zz}^{Hz}(z',z|\vec{k}) & = & \frac{e^{-k(z+z')}}{2 \gamma^{2}}
\{ \gamma^{2}+(\gamma^{2}-1)kz +(\gamma^{2}+1)kz' + 2(\gamma^{2}-1)
k^{2}zz' \} \nonumber \\
\sigma_{z\beta}^{Hz}(z',z|\vec{k}) & = & -\frac{ik_{\beta}}{2 \gamma^{2}k}
e^{-k(z+z')}
\{ 1+(\gamma^{2}-1)kz -(\gamma^{2}+1)kz' - 2(\gamma^{2}-1)
k^{2}zz' \} \nonumber \\
\sigma_{\alpha \beta}^{Hz}(z',z|\vec{k}) & = & \frac{e^{-k(z+z')}}{2 
\gamma^{2}(\gamma^{2}-1)k^{2}}
\{ (\gamma^{2}-2)[(\gamma^{2}+1)+2(\gamma^{2}-1)kz ]q_{\alpha}q_{\beta}
\nonumber \\ & & +(\gamma^{2}-1)[(\gamma^{2}+2) + 3(\gamma^{2}-1)kz
-(\gamma^{2}+1)kz' - 2(\gamma^{2}-1)
k^{2}zz']k_{\alpha}k_{\beta} \} \nonumber \\
\sigma_{zz}^{H\delta}(z',z|\vec{k}) & = & \frac{i k_{\delta}}{2 \gamma^{2}k}
e^{-k(z+z')}
\{ 1-(\gamma^{2}-1)kz +(\gamma^{2}+1)kz' - 2(\gamma^{2}-1)
k^{2}zz' \} \nonumber \\
\sigma_{\beta z}^{H\delta}(z',z|\vec{k}) 
& = & \frac{e^{-k(z+z')}}{2 \gamma^{2}k^{2}}
\{ \gamma^{2} q_{\delta}q_{\beta} +[\gamma^{2}
-(\gamma^{2}-1)kz -(\gamma^{2}+1)kz' + 2(\gamma^{2}-1)
k^{2}zz']k_{\delta} k_{\beta} \} \nonumber \\
\sigma_{\alpha \beta}^{H\delta}(z',z|\vec{k}) & = & 
\frac{i e^{-k(z+z')}}{2 \gamma^{2}(\gamma^{2}-1)k^{3}}
\{ (\gamma^{2}-2)[(\gamma^{2}+1)-2(\gamma^{2}-1)kz]
k_{\delta} q_{\alpha}q_{\beta} + 
\nonumber \\ & & 
(\gamma^{2}-1)[2\gamma^{2}+1-3(\gamma^{2}-1)kz 
-(\gamma^{2}+1)kz' + 2(\gamma^{2}-1)
k^{2}zz']k_{\delta}k_{\alpha}k_{\beta} \} \nonumber \\
& & + \gamma^{2}(\gamma^{2}-1)
q_{\delta} [k_{\alpha}q_{\beta}+k_{\beta}q_{\alpha}] \} \: .
\label{shf}
\eea
Finally, we derive the form of the Fourier components of the
``vector potential'' $A_{\beta}^{zzz}(z',z|\vec{k})$ and
$A_{\beta}^{zz \delta}(z',z|\vec{k})$, necessary for the 
integration of Eq. (\ref{sze2}). From Eqs. (\ref{adef}) and
(\ref{sfth}):
\beq
a_{\beta}^{zz}(z',z|\vec{k}) = -iq_{\eta}\chi _{\eta \beta}^{z}
=-\frac{iq_{\beta}e^{-k(z+z')}}{2\gamma^{2}k^{2}} [ \gamma^{2}
+(\gamma^{2}-1)kz +(\gamma^{2}+1)kz'+2(\gamma^{2}-1)k^{2}zz'] \: ,
\eeq
and from Eq. (\ref{aa}):
\beq
A_{\beta}^{zzz}(z',z|\vec{k}) =-( \frac{\partial}{\partial z}
+\frac{\partial}{\partial z'})a_{\beta}^{zz}(z',z|\vec{k})=
-\frac{2iq_{\beta}}{\gamma^{2}k}kz'
e^{-k(z+z')}[1+(\gamma^{2}-1)kz] \: ,
\eeq
or $A_{\beta}^{zzz}(z'=0,z|\vec{k})=0$. Similarly:
\bea
a_{\beta}^{z\delta}(z',z|\vec{k}) & = & -iq_{\eta}\chi _{\eta 
\beta}^{\delta}
=\frac{e^{-k(z+z')}}{2\gamma^{2}k^{3}} \{ \gamma^{2} q_{\delta}k_{\beta}
\nonumber \\ & & + [ 1
-(\gamma^{2}-1)kz +(\gamma^{2}+1)kz'-2(\gamma^{2}-1)k^{2}zz'] 
k_{\delta}q_{\beta} \} \: ,
\eea
and from Eq. (\ref{aa}):
\beq
A_{\beta}^{zz\delta}(z',z|\vec{k}) =-( \frac{\partial}{\partial z}
+\frac{\partial}{\partial z'})a_{\beta}^{z\delta}(z',z|\vec{k})=
\frac{e^{-k(z+z')}}{\gamma^{2}k^{2}}\{ \gamma^{2}k_{\beta}
q_{\delta}+2[\gamma^{2}-(\gamma^{2}-1)kz]kz'k_{\delta}
q_{\beta} \} \: ,
\eeq
or $A_{\beta}^{zz\delta}(z'=0,z|\vec{k}) = \exp (-kz) k_{\beta}
q_{\delta}/k^{2}$.

\section*{Appendix II: Green's function of a thin plate}
We write the Fourier components of the surface Green's function appropriate
 for a thin plate, $g_{ij}^{(p)}(z|\vec{k})$ (they are the solution to
Eqs. (\ref{egp}) and (\ref{bgp})):
\bea
g_{zz}^{(p)}(z,\xi|\vec{k}) & = & \frac{1}{4 \mu (\gamma^{2}-1)k}
\{\gamma^{2}\phi_{2}(z|\xi)-(\gamma^{2}-1)k z \phi_{3}(z|\xi)
+(\gamma^{2}-1)k \xi \phi_{4}(z|\xi) \} \nonumber \\
g_{z \beta}^{(p)}(z,\xi|\vec{k}) & = & \frac{i k_{\beta}}
{4 \mu (\gamma^{2}-1)k^{2}}
\{(\gamma^{2}-1)kz \phi_{1}(z|\xi)-
(\gamma^{2}-1)k \xi \phi_{2}(z|\xi) -\phi_{4}(z|\xi) \} \nonumber \\
g_{\beta z}^{(p)}(z,\xi|\vec{k}) & = & -\frac{i k_{\beta}}
{4 \mu (\gamma^{2}-1)k^{2}}
\{ (\gamma^{2}-1)k \xi \phi_{1}(z|\xi) - 
(\gamma^{2}-1)k z \phi_{2}(z|\xi) - \phi_{3}(z|\xi) \} \nonumber \\
g_{\alpha \beta}^{(p)}(z,\xi|\vec{k}) & = & \frac{k_{\alpha}k_{\beta}}
{4 \mu (\gamma^{2}-1)k^{3}}
\{ (\gamma^{2}-1)k z \phi_{4}(z|\xi) - (\gamma^{2}-1) k\xi \phi_{3}(z|\xi)
+\gamma^{2} \phi_{1}(z|\xi) \} \nonumber \\
& + &  (\pm) \frac{q_{\alpha}q_{\beta}}{2 \mu k^{3}} 
[ \frac{\cosh (kz)}{\sinh (k \xi )} +
 \frac{ \sinh (k z) }{\cosh (k \xi) } ] \: ,
\eea
where $\xi = \pm h$, and the functions $\phi_{1}(z|\xi)$ to 
$\phi_{4}(z|\xi)$ are defined
in the following way:
\bea
\phi_{1}(z|\xi) & = & \frac{1}{\delta_{s}} \cosh(k z) \cosh (k \xi)
+ \frac{1}{\delta_{a}} \sinh(kz) \sinh (k \xi) \nonumber \\
\phi_{2}(z|\xi) & = & \frac{1}{\delta_{s}} \sinh(k z) \sinh (k \xi)
+ \frac{1}{\delta_{a}} \cosh(kz) \cosh (k \xi) \nonumber \\
\phi_{3}(z|\xi) & = & \frac{1}{\delta_{s}} \cosh(k z) \sinh (k \xi)
+ \frac{1}{\delta_{a}} \sinh(kz) \cosh (k \xi) \nonumber \\
\phi_{4}(z|\xi) & = & \frac{1}{\delta_{s}} \sinh(k z) \cosh (k \xi)
+ \frac{1}{\delta_{a}} \cosh(kz) \sinh (k \xi)  \: ,
\eea
with:
\bea
\delta_{s} \equiv \sinh (kh) \cosh (kh) + kh \nonumber \\
\delta_{a} \equiv \sinh (kh) \cosh (kh) - kh \: .
\eea
Doing the calculation in Eq. (\ref{shp}), one gets the Fourier components
$\sigma_{ij}^{H m}(z',z|\vec{k})$:
\bea
\sigma_{zz}^{Hz}(z',z|\vec{k}) & = & \sum_{\xi} (\pm) 
\frac{e^{-k|\xi -z|}}{4 \gamma^{2}} \{ 
(1+(\gamma^{2}-1)k|\xi-z|)(kz' \phi_{4}(z'|\xi)-k\xi \phi_{3})
\nonumber \\ & & -
s(\xi-z)(\gamma^{2}+(\gamma^{2}-1)k|\xi-z|)
(\phi_{3}+k\xi \phi_{1}-kz' \phi_{2})
 \} \nonumber \\
\sigma_{z\beta}^{Hz}(z',z|\vec{k}) & = & -\sum_{\xi} (\pm) 
\frac{i k_{\beta}e^{-k|\xi -z|}}{4 \gamma^{2}k} \{ 
s(\xi-z)(\gamma^{2}+(\gamma^{2}-1)k|\xi-z|)
(kz' \phi_{3}(z'|\xi)-k\xi \phi_{4})
\nonumber \\ & & +(1+
(\gamma^{2}-1)k|\xi-z|)(\phi_{4}+kz' \phi_{1}-k\xi \phi_{2}) \} \nonumber \\
\sigma_{\alpha \beta}^{Hz}(z',z|\vec{k}) & = & -\sum_{\xi} (\pm) 
\frac{e^{-k|\xi -z|}}{4 \gamma^{2}(\gamma^{2}-1)} \{ 
q_{\alpha}q_{\beta}(\gamma^{2}-2) \times
\nonumber \\ & & [s(\xi-z)(\gamma^{2}+
(\gamma^{2}-1)k|\xi-z|) \phi_{3}
+ (1+(\gamma^{2}-1)k|\xi-z|) \phi_{1}]
\nonumber \\ & &
+ k_{\alpha}k_{\beta} (\gamma^{2}-1) [ s(\xi -z) (
\gamma^{2}+(\gamma^{2}-1)k|\xi -z|)(\phi_{3}+kz' \phi_{2}-
k \xi \phi_{1}) 
\nonumber \\ & & + (1+(\gamma^{2}-1)k|\xi-z|)(2 \phi_{1}+
kz' \phi_{4} -k \xi \phi_{3}) ]
 \} \nonumber \\
\sigma_{zz}^{H\delta}(z',z|\vec{k}) & = & \sum_{\xi} (\pm) 
\frac{ik_{\delta}e^{-k|\xi -z|}}{4 \gamma^{2}k} \{ 
(1-(\gamma^{2}-1)k|\xi-z|)(\phi_{3}+k\xi \phi_{1}-kz' \phi_{2})
\nonumber \\ & & 
-s(\xi-z)(kz'\phi_{4}-k\xi \phi_{3})(\gamma^{2}-(\gamma^{2}-1)k
|\xi-z|)
 \} \nonumber \\
\sigma_{z\beta}^{H\delta}(z',z|\vec{k}) & = & \sum_{\xi} (\pm) 
\frac{e^{-k|\xi -z|}}{4 \gamma^{2}k^{2}} \{ k_{\delta}k_{\beta}[ 
(1-(\gamma^{2}-1)k|\xi-z|)(k \xi \phi_{4}-kz' \phi_{3})
\nonumber \\ & & 
-s(\xi-z)(\phi_{4}+kz'\phi_{1}-k\xi \phi_{2})(\gamma^{2}-(\gamma^{2}-1)k
|\xi-z|)] \nonumber \\
&  & -(\pm)q_{\delta}q_{\beta} s(\xi-z) \gamma^{2}[\frac{\sinh(kz')}
{\sinh(k\xi)}+\frac{\cosh(kz')}{\cosh(k\xi)} ] \} \nonumber \\
\sigma_{\alpha \beta}^{H\delta}(z',z|\vec{k}) & = & \sum_{\xi} (\pm) 
\frac{e^{-k|\xi -z|}}{4 \gamma^{2}(\gamma^{2}-1)k^{3}} \{ 
ik_{\delta}k_{\alpha}k_{\beta}(\gamma^{2}-1) \times \nonumber \\  
& & 
[(1-(\gamma^{2}-1)k|\xi-z|)(\phi_{3}+k z' \phi_{2}-k\xi \phi_{1}) +
\nonumber \\ & & 
s(\xi-z) (\gamma^{2}-(\gamma^{2}-1)k|\xi-z|)
(2\phi_{1}+kz'\phi_{4}-k\xi \phi_{3})] +
i k_{\delta} q_{\alpha}q_{\beta} (\gamma^{2}-2)
\times \nonumber \\ & & [
(1-(\gamma^{2}-1)k|\xi-z|)\phi_{3}+s(\xi-z)(\gamma^{2}-
(\gamma^{2}-1)k|\xi-z|)\phi_{1}] 
\nonumber \\ & & + (\pm)i
s(\xi-z) \gamma^{2}(\gamma^{2}-1)[\frac{\cosh(kz')}
{\sinh(k\xi)}+\frac{\sinh(kz')}{\cosh(k\xi)}]q_{\delta} [
k_{\alpha}q_{\beta}+k_{\beta}q_{\alpha} ] 
 \} \: ,
\nonumber \\
\eea
with $s(\xi-z) \equiv sign(\xi-z)$.
Finally, we write the Fourier components of the
``vector potential'' evaluated at the free surfaces, necessary for the
integration of Eq. (\ref{szp2}):
\bea
 A_{\beta}^{zzz}(z'=\pm h,z|\vec{k}) & = & 0 \nonumber \\
A_{\beta}^{zz\delta}(z'=\pm h,z|\vec{k}) & = &
-\frac{k_{\beta}q_{\delta}}{2k^{2}} \{ \pm \frac{\cosh (kz)}
{\cosh (kh)} + \frac{\sinh (kz)}{\sinh (kh)} \} \: .
\eea

\section*{References}
 
\begin{description}

\item Boudet, J.F., Ciliberto S., and Steinberg J.
(1996) Dynamics of crack propagation in brittle materials.
 {\it J. Phys. II France}, {\bf 6}, 1493.

\item Chishko K.A. (1989) Dynamical Green's tensor and elastic fields
of a system of moving dislocation loops in an isotropic plate.
{\it  Sov. Phys. Acoust.} {\bf 35}, 307.

\item Eshelby, J.D., and Stroh A.N. (1951)
Dislocations in thin plates. {\it Philos. Mag.}, 
{\bf 42}, 1401.

\item Gosling, T.J., and Willis, J.R. (1994) A line integral 
representation for the stresses due to an arbitrary dislocation in
a half space. {\it J. Mech. Phys. Solids},
{\bf  42}, 1199.

\item Lund, F. (1988) Response of a stringlike dislocation loop to
an external stress. {\it J. Mater. Res.}, {\bf 3}, 280.

\item Lund, F. (1996) Elastic forces that do no work and the dynamics
of fast cracks. {\it Phys. Rev. Lett.}, {\bf 76}, 2742.

\item Mura, T. (1963) Continuous distribution of moving
dislocations. {\it Philos. Mag.} {\bf 8}, 843.

\item Poruchikov V.B. (1993) {\it Methods of the classical theory of
elastodynamics}. Springer Verlag.

\item Sharon, E., Gross, S.P, and Fineberg, J. (1996)
Energy dissipation in dynamic fracture.
{\it Phys. Rev. Lett.}, {\bf 76}, 2117.

\end{description}

\end{document}